\documentstyle[12pt,epsfig]{article}
	\newdimen\eqskip
	\newdimen\txtskip
	\baselineskip=\txtskip
\begin{document}

  \newcommand{\ccaption}[2]{
    \begin{center}
    \parbox{0.85\textwidth}{
      \caption[#1]{\small{{#2}}}
      }
    \end{center}
    }
\def\vector#1{{\bf#1}}
\def\sigtot{\sigma_{\rm tot}}
\def\half{\frac{1}{2}}
\def\theonetwo{\theta_{12}}
\def\thethrfour{\theta_{34}}
\newcommand\sss{\scriptscriptstyle}
\newcommand\mq{\mbox{$m_{\sss \rm Q}$}}
\newcommand\mug{\mu_\gamma}
\newcommand\mue{\mu_e}
\newcommand\muf{\mbox{$\mu_{\sss F}$}}
\newcommand\mur{\mbox{$\mu_{\sss R}$}}
\newcommand\muo{\mbox{$\mu_0$}}
\newcommand\ep{\epsilon}
\newcommand\avr[1]{\left\langle #1 \right\rangle}
\newcommand\lambdamsb{\mbox{$
\Lambda_5^{\rm \sss \overline{MS}}
$}}
\def	\be		{\begin{equation}}
\def	\ee		{\end{equation}}
\def	\ba		{\begin{eqnarray}}
\def	\ea		{\end{eqnarray}}
\def	\nn		{\nonumber}
\def	\=		{\;=\;}
\def	\frac		#1#2{{#1 \over #2}}
\def	\ret		{\\[\eqskip]}
\def	\to		{\rightarrow }
\def	\ie		{{\em i.e.\/} }
\def	\eg		{{\em e.g.\/} }
\def	\z		{\mbox{$Z$}}
\def	\w		{\mbox{$W$}}
\def	\g		{\mbox{$\gamma$}}
\def	\e		{\mbox{$e$}}
\def	\m		{\mbox{$\mu$}}
\def	\b		{\mbox{$b$}}
\def	\bbar		{\mbox{$\bar b$}}
\def 	\ttbar 		{\mbox{$t \bar t$}}
\def	\jpsi		{\mbox{$\psi$}}
\def	\psip		{\mbox{$\psi^\prime$}}
\def\vevpsp{\mbox{$\langle {\cal O}_8^{\psi'}(^3S_1) \rangle$}}
\def    \pt		{\mbox{$p_{\sss \rm T}$}}
\def    \ptmin		{\mbox{$p_{\sss \rm T}^{min}$}}
\def \ptpair {\mbox{$p^{t \bar t}_{\scriptscriptstyle T}$}}
\def    \et		{\mbox{$E_{\sss \rm T}$}}
\def	\sumet		{$\sum E_T$}
\def	\mjj		{\mbox{$M_{jj}$}}
\def	\as		{\mbox{$\alpha_s$}}
\def	\dphi		{\mbox{$\Delta\phi$}}
\def	\scnd		{2$^{nd}$}
\def	\trd		{3$^{rd}$}
\begin{titlepage}
\nopagebreak
{\flushright{
        \begin{minipage}{5cm}
        CERN-TH/95-191\\
	hep-ph/9508260 \\
        \end{minipage}        }

}
\vfill
\begin{center}
{\LARGE { \bf \sc
Heavy Quark Production \\[0.5cm]
In Hadronic Collisions} }
\footnote{To appear in the Proceedings of the 6th International Symposium on
Heavy Flavour Physics, Pisa, Italy, June 6-10, 1995.}
\vfill
\vskip .5cm
{\bf Michelangelo L. MANGANO}
\footnote{On leave of absence from INFN, Pisa, Italy}
\vskip .3cm
{CERN, TH Division, Geneva, Switzerland} \\
{E-mail: mlm@vxcern.cern.ch}
\end{center}
\nopagebreak
\vfill
\begin{abstract}
We review the physics of heavy quark and quarkonium production in high energy
hadronic collisions. We  discuss the status of the theoretical calculations and
compare the current results with the most recent measurements from the Tevatron
collider   experiments.
\end{abstract}
\vskip 1cm
CERN-TH/95-191\hfill \\
July 1995 \hfill
\vfill
\end{titlepage}

\section{Introduction}
Heavy quark production in high energy hadronic collisions constitutes a
benchmark process for the study of perturbative QCD and an important tool to
explore flavour physics.
\begin{itemize}
\item $b$ quarks are produced in abundance in hadronic
collisions, and will eventually allow ultimate tests of CP
violation in the $b$ system, as well as studies of rare $b$ decays with
branching ratio levels of the order of $10^{-10}$. A
detailed understanding of the
production properties at future machines (such as the LHC) is therefore of
the utmost importance.
\item The comparison of the current experimental data with the predictions of
QCD provides a necessary check that the ingredients entering the evaluation of
hadronic processes (partonic distribution functions and higher order
corrections) are under control and can be used to evaluate the rates for more
exotic phenomena or to extrapolate the calculations to even higher energies.
\item Accurate studies of the production properties of the $top$ quark
rely on a solid understanding of the QCD dynamics, in order to
isolate possible contributions from new phenomena.
\item Measurements of heavy quark production in fixed target experiments are
dominated by data at low \pt, a region where non-perturbative effects are not
negligible. The comparison of data with the expectations of perturbative QCD
offers the possibility to explore some interesting features of
non-perturbative hadronic dynamics \cite{fmnr}.
\item At HERA, $c$ and $b$ quarks are largely produced via photon-gluon fusion,
therefore providing a direct probe on the gluon density of the proton,
complementary to the information extracted from the measurement of structure
functions. First data have already become available, and have been shown at
this Conference.
\item Production of quarkonium states, in addition to providing yet another
interesting framework for the study of the boundary between perturbative and
non-perturbative QCD, is important in view of the use of exclusive
charmonium decays of $b$ hadrons for the detection of CP violation phenomena.
\end{itemize}
In this presentation we review the current status of theoretical
calculations, and discuss the implications of the most recent experimental
measurements of $b$ quarks and charmonium states
performed at the Tevatron $p\bar p$ collider.

\section{Open Flavour Production: Theory Overview}
To start with, we briefly report on the current status of the theoretical
calculations. A distinction must be made between calculations performed at a
complete but fixed order in perturbation theory (PT), and those performed by
resumming classes of potentially large logarithmic contributions that arise at
any order in PT. The exact matrix elements squared for heavy quark production
in hadronic collisions are fully known up to the ${\cal O}(\as^3)$, for both
real and virtual processes. These matrix elements have been used to evaluate at
the next-to-leading order (NLO) the total production cross section \cite{btot},
single-particle inclusive distributions \cite{bpt}\ and two-particle inclusive
distributions (a.k.a. correlations) \cite{mnr}.

Three classes of large logarithms can appear in the perturbative expansion for
heavy quark production:
\begin{enumerate}
\item $[\as\log(S/m_Q^2)]^n \sim [\as\log(1/x_{\rm Bj})]^n$ terms, where $S$ is
the hadronic CM energy squared. These small-$x$ effects are possibly relevant
for  the production of charm or bottom quarks at the current energies, while
they should have no effect on the determination of the top quark cross section,
given the large $t$ mass. Several theoretical studies have been performed
\cite{smallx}, and the indications are that \b\ production cross sections
should not increase by more than 30--50\% at the Tevatron energy because of
these effects.
\item $[\as\log(m_Q/p^T_{QQ})]^n$ terms, where $p^T_{QQ}$ is the transverse
momentum of the heavy quark pair. These contributions come from the multiple
emission of initial-state soft gluons, similarly to standard Drell Yan
corrections. These corrections have been studied in detail in the case of top
production, where the effect is potentially large due to the heavy top mass
\cite{laenen}. They are not relevant to the total cross
section of \b\ quarks, but affect the kinematical distributions of pairs
produced just above threshold \cite{berger2}, or in regions at the edge of
phase space, such as $\Delta\phi=\pi$.
\item $[\as\log(p_T/m_Q)]^n$ terms, where $p_T$ is the transverse momentum of
the heavy quark. These terms arise from multiple collinear gluons emitted by
a heavy quark produced at large transverse momentum, or from the
branching of gluons into heavy quark pairs. Again these corrections are
not expected to affect the total production rates, but will contribute to the
large-\pt\ distributions of $c$ and $b$ quarks. No effect is expected for the
top at current energies. These logarithms can be resummed using a fragmentation
function formalism \cite{greco},
with a significant improvement in the stability w.r.t. scale changes
for $\pt>50$ GeV.
\end{enumerate}

\section{Single Inclusive Bottom Production}
\begin{figure}[t]
\centerline{\psfig{figure=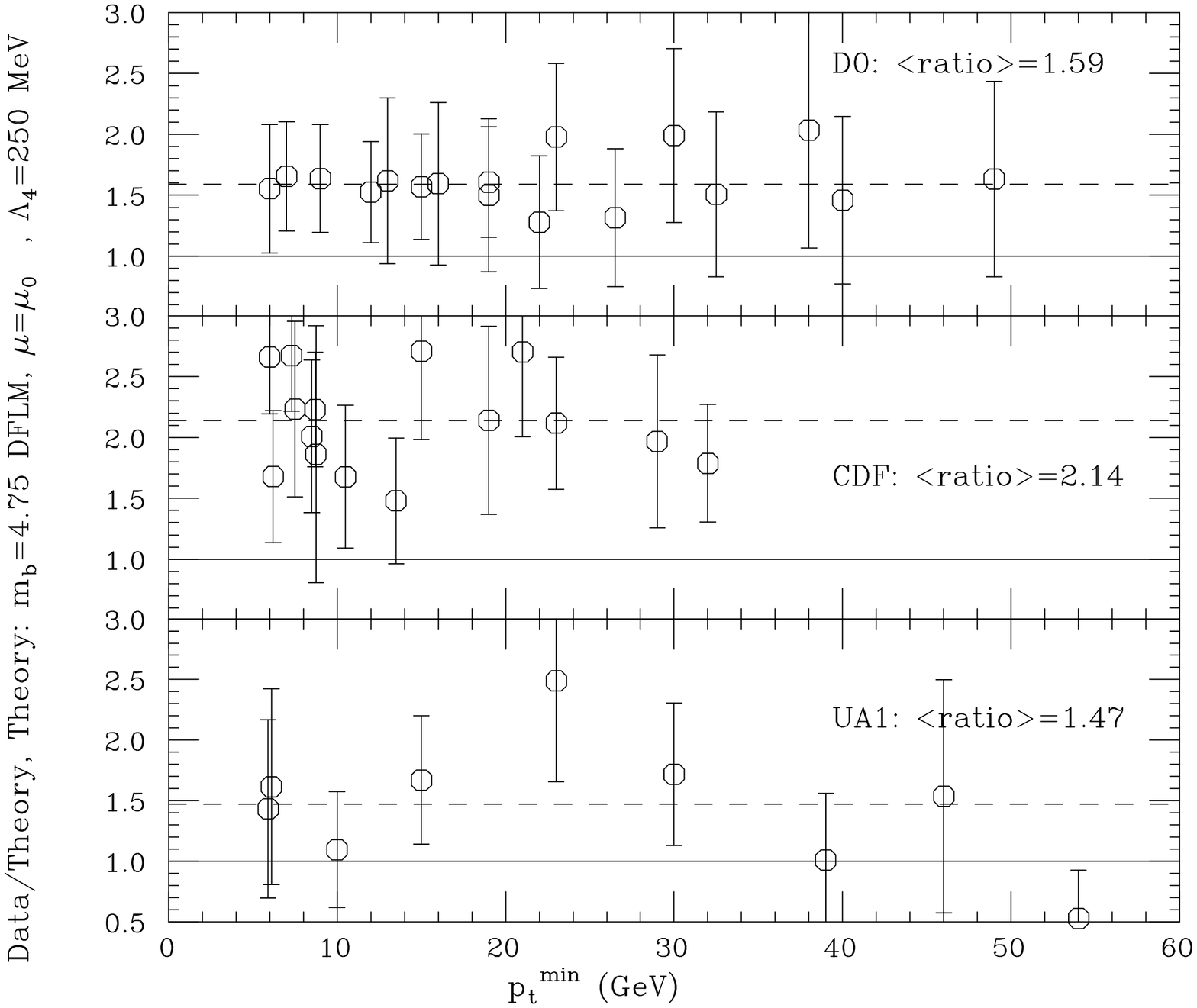,width=10cm,angle=90}}
\ccaption{}{Ratio of data and theory for the integrated $b$ \pt\ distribution
at
UA1, CDF and D0. Theory evaluated at NLO \cite{bpt}\ using DFLM parton
densities \cite{dflm}\ ($\lambdamsb=173$ MeV), $m_b=4.75$ GeV,
$\muf=\mur=\sqrt{m^2+p_T^2}$. \label{ratio-dflm}}
\end{figure}

The status of $b$ production at hadron colliders has been quite puzzling for
some time. Data collected by UA1~\cite{ua1b} at the CERN
S$p\bar p$S collider ($\sqrt S$ = 630
GeV) were in good agreement with theoretical expectations based on the NLO QCD
calculations \cite{bpt}.
On the contrary, the first measurements performed at 1.8
TeV by the CDF \cite{cdfpre93} experiment at the Fermilab collider showed a
significant discrepancy with the same calculation.
The new data presented at this Conference \cite{paulini}\
by the two Fermilab experiments, CDF and D0, allow us to draw a more complete
picture of the situation.

\begin{figure}[t]
\centerline{\psfig{figure=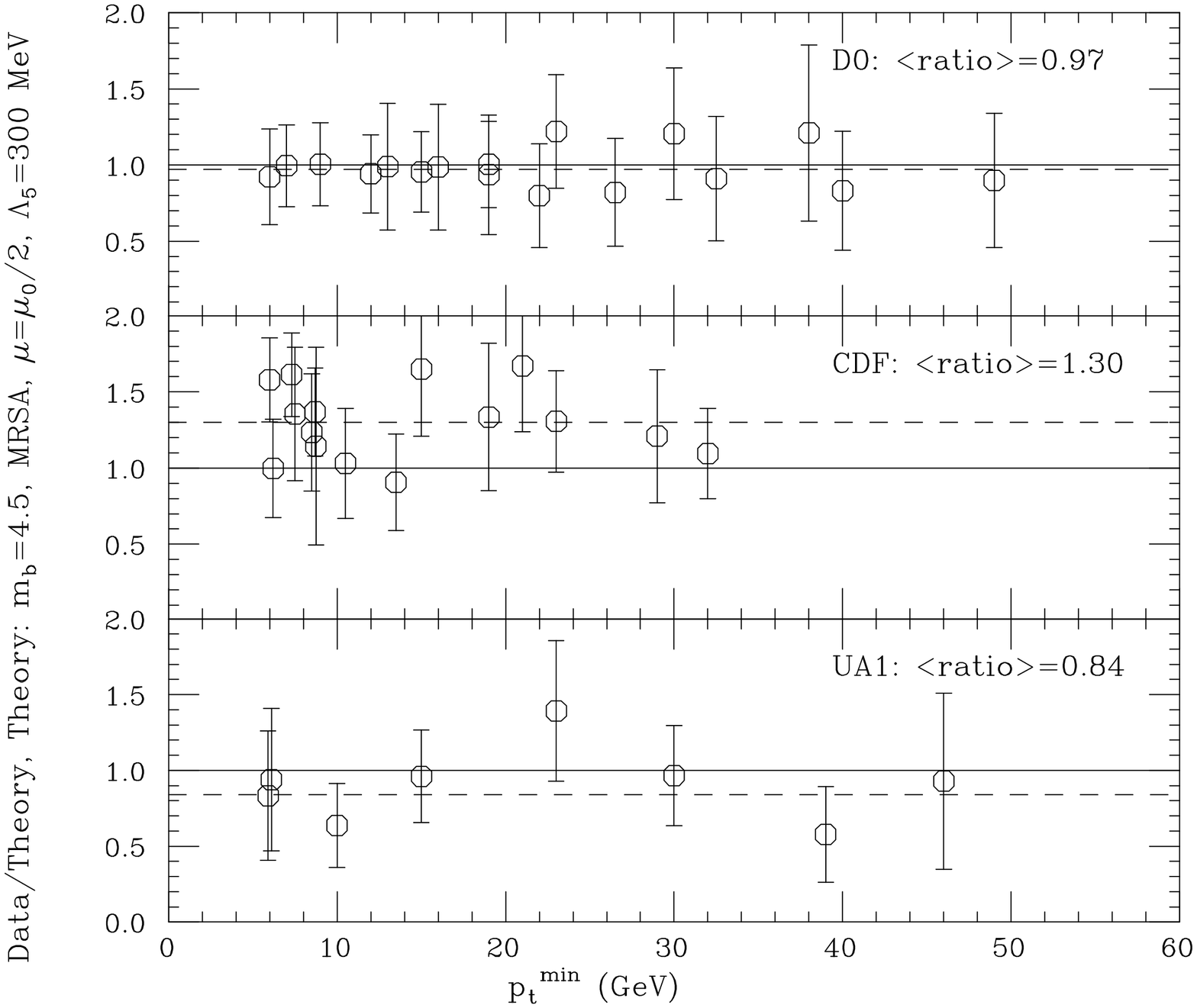,width=10cm,angle=90}}
\ccaption{}{Ratio of data and theory for the integrated $b$ \pt\ distribution
at
UA1, CDF and D0. Theory evaluated at NLO \cite{bpt}\ using MRSA parton
densities \cite{mrsa}, $\lambdamsb = 300$ MeV, $m_b = 4.5$ GeV,
\muf = \mur = $\sqrt{m^2+p_T^2}/2$. \label{ratio-mrsa}}
\end{figure}
We present all three sets of data from UA1, CDF and D0 in a single
plot, containing the ratio of the measurements to the theory, for a uniform
choice of parameters entering the theoretical calculation. In
fig.~\ref{ratio-dflm}, we choose the same theoretical prediction as was
available at
the time of the UA1 measurements, namely the central set of the DFLM
\cite{dflm}\ parton distributions ($\lambdamsb=173$ MeV), $m_b=4.75$ GeV and
renormalization/factorization scales equal to the transverse mass of the $b$
quark, $\muf=\mur=\sqrt{m^2+p_T^2}\equiv \muo $. Depending on whether we use
the D0 or the CDF data as representative of the   $b$ cross section at 1.8 TeV,
we can draw two different conclusions. The plot shows clearly that the ratio
{\em data/theory} is the same, at UA1 and at D0,
to within less than 10\%.
While larger than 1, this ratio can be reduced by selecting different input
parameters, still in the range of acceptable values. For example,
fig.~\ref{ratio-mrsa}\ shows the same distributions with the theory
evaluated using the more recent set of parton densities MRSA \cite{mrsa},
$m_b=4.5$ GeV, $\muf=\mur=\muo/2$ and $\lambdamsb=300$ MeV, a value
close to the LEP measurement of \as. With this choice of parameters the
agreement between NLO QCD and data is perfect, both at 630 and at 1800 GeV. The
data from CDF are in good agreement with the theory shape, but are about
30--40\% higher in
normalization relative to the D0 ones. If one were to choose CDF data as
representative of the Tevatron rate, the conclusion would be that the $b$
production cross section grows between 630 and 1800 GeV by a factor of 40\%
more than NLO QCD predicts. An effect of such a size would be in agreement with
the evaluation of the small-$x$ effects mentioned earlier. The only conclusions
we can therefore draw from the current data are that:
\begin{enumerate}
\item the comparison of data and NLO predictions for the $b$ production at 630
and 1800 GeV favours small values of the $b$ mass and values of \as\ consistent
with the LEP measurement;
\item the relative difference in the data/theory ratio at 630 and 1800~GeV is
at most 40\%, value obtained using the CDF data. This is consistent with the
estimated effect of small-$x$ higher order corrections;
\begin{figure}[t]
\begin{center}
\psfig{figure=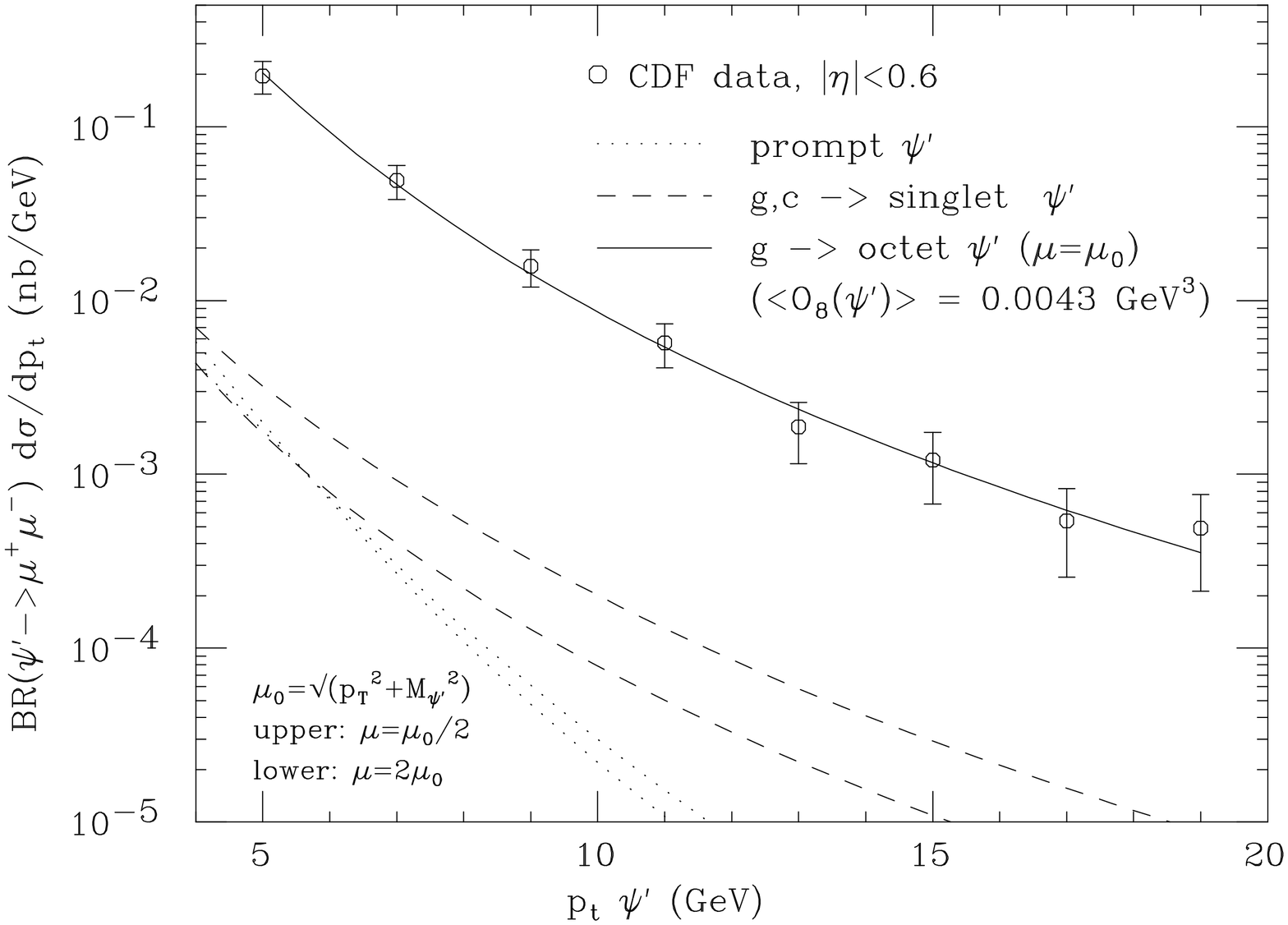,width=10cm,angle=90,clip=}
\ccaption{}{\label{fpspdata}\small Inclusive prompt $\psip$ $p_T$ distribution.
CDF data versus theory. We show the contribution of the different sources.
Dotted lines: LO production in the CSM; dashed lines: gluon and charm
fragmentation in the CSM; solid line: gluon fragmentation in the colour octet
mechanism.}
\end{center}
\end{figure}
\item there is a residual 30--40\% discrepancy in absolute normalization
between
the CDF and the D0 results, that will need to be resolved before additional
progress can be made in interpreting the data.
\end{enumerate}

\section{Charmonium production}
The production of heavy quarkonium states in high energy processes has recently
attracted a lot of theoretical and experimental interest.
Detailed measurements of differential cross sections for
production of \jpsi, \psip\ and $\chi$ states have recently become available
[14--18], and have been reviewed at this
Conference \cite{sansoni}.
Theoretical models for production have existed for several years (see
ref.~\cite{schuler} for a comprehensive review and references).
The comparison of these models with the most recent data
has shown dramatic discrepancies, the most striking one (theory predicting
a factor of 50 fewer prompt \psip\ than measured by CDF
\cite{cdf94}) having become known as the ``CDF anomaly''.
Attempts to explain the features of these data have recently led
to a deeper theoretical understanding of the mechanisms of
quarkonium production. I will briefly summarize here this progress (for a more
complete review, see ref.~\cite{mlmppbar}).

\begin{figure}[t]
\begin{center}
\psfig{figure=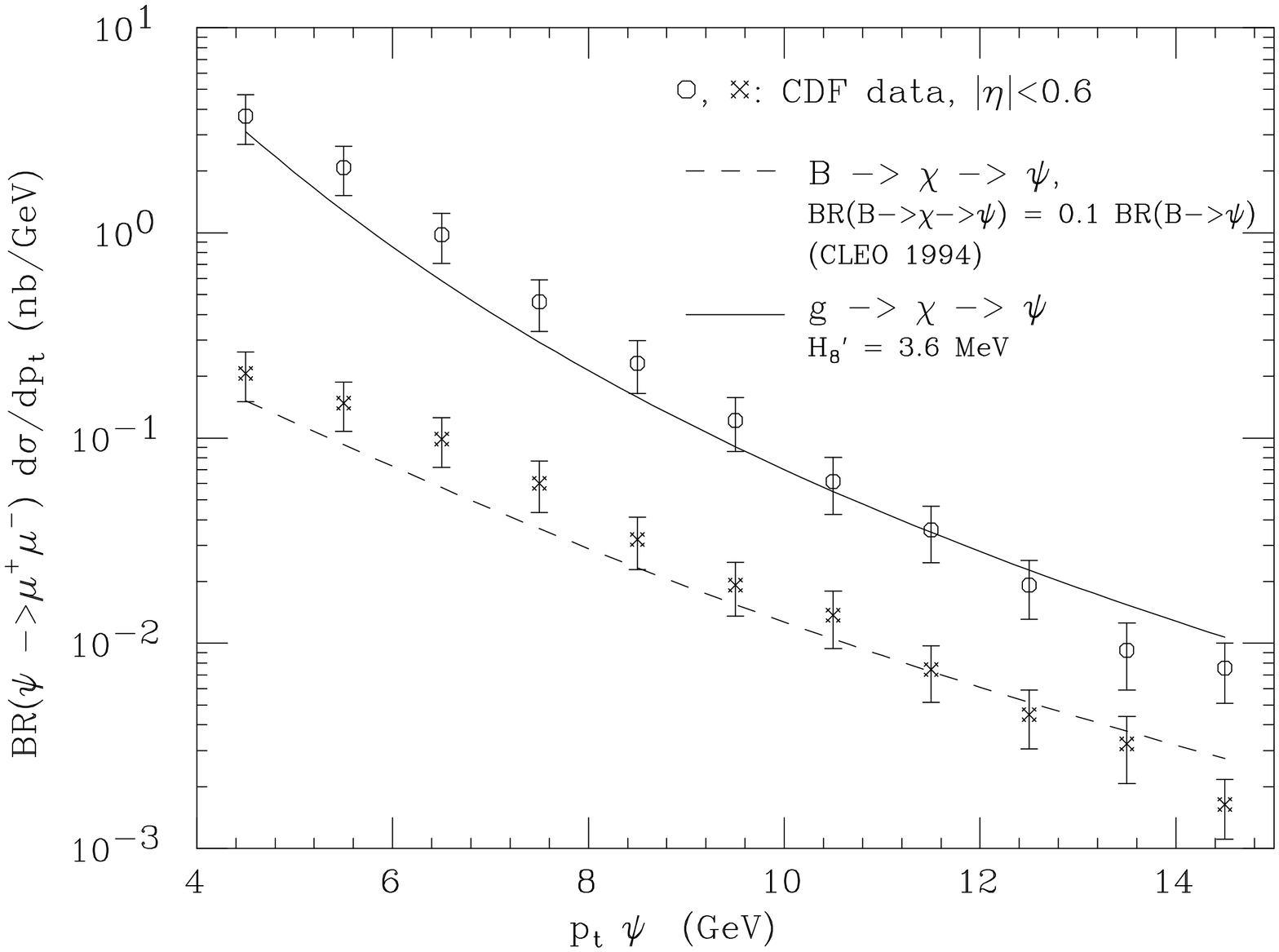,width=10cm,angle=90,clip=}
\ccaption{}{
\label{fchi}\small Inclusive $p_T$ distribution of $\jpsi$'s from
$\chi_c$ production and decay.
}
\end{center}
\end{figure}
The production of quarkonium at large \pt\ is a phenomenon with two
different time scales: the shorter time scale corresponds to the generation of
the heavy quark pair, the longer one to the binding of the pair.
In all models, it is assumed that the details of the bound state
formation can be absorbed into some non-perturbative parameter, directly
related to the value of the non-relativistic wave function at the origin. Where
the models differ is in specifying how the heavy quark pair, produced by the
hard scattering in a generic colour and angular momentum state, evolves into a
state with the right quantum numbers to form the desired hadron.
In the first QCD-based model, the so-called colour singlet model (CSM,
\cite{csm}), it was assumed that the quark pair is produced in a colour singlet
state with the right angular momentum already during the hard collision. It is
easy to show that the production of quarkonium in the CSM
is heavily suppressed at large \pt, mainly because it is difficult to hold the
bound state together when this is probed at the large virtualities typical of a
high-\pt\ phenomenon \cite{mlmppbar}. In other words, in the naive CSM,
production of quarkonium is a higher-twist effect, highly suppressed by a
form-factor-like damping as soon as \pt\ becomes larger than the mass of
the state.

This prediction has been recently disproved by the data.
The dotted line in fig.~\ref{fpspdata}, for example, shows the prediction of
the
CSM model for \psip\ production at the Tevatron, compared to the CDF data
\cite{cdf94}. Not only is the overall normalization of the theory curve
significantly lower than the data, but also the shape is much steeper than
observed.

It has been pointed out recently \cite{bygluon}\  that higher order
contributions in \as\ can dominate production at large \pt.
The process responsible for these contributions is the splitting of a
large-\pt\ gluon into a $Q\bar Q$ pair, which then evolves into a colour
singlet
state by emission of one or more perturbative gluons. The additional powers of
\as\ required for this process are largely compensated by the absence of a
form factor suppression. It turns out, in fact, that these terms are of order
$[\as \times (\pt/m)^2]^n$ relative to the LO diagrams ($n=1$ in the case of
$\chi$ production, and $n=2$ in the case of \jpsi\ or \psip), and become
dominant as soon as \pt\ is slightly larger than $m$, the quarkonium mass.

The effect of these {\em fragmentation} contributions is shown by the dashed
line in fig.~\ref{fpspdata}: the \pt\ shape is now correct, although the total
rate is still low by more than an order of magnitude. A similar behaviour is
observed in the production of \jpsi's. On the contrary, the
predictions for $\chi$ production, as shown in fig.~\ref{fchi}, agree with
the data both in shape and in rate \cite{pheno}.

A possible solution to this remaining discrepancy in the \jpsi\ and \psip\
sector was recently suggested by Braaten and Fleming \cite{bf}. Their proposal
is based on the observation that, contrary to the case of the gluon
fragmentation into \jpsi\ states, where the emitted gluons are both hard,
the fragmentation process into $\chi$ states is dominated by emission of a soft
gluon. A large infrared logarithm enhances the fragmentation rate of a gluon
into $\chi$'s relative to that into \jpsi's.
This logarithm signals the presence in the $\chi$ state
of a  large component made by a $c \bar c$ pair in a colour octet $^3S_1$
state, accompanied by an on-shell gluon \cite{bbl}. This component has a
non-zero overlap with the $c\bar c$ state produced by the splitting of the
large-\pt\ gluon. Braaten and Fleming suggested that a similar
colour octet $^3S_1$ component might be present in the relativistic expansion
of the \jpsi\ and \psip\ wave function. The work of ref.~\cite{bbl}\ indicates
that such a component should have an amplitude of order $v^2$ relative
to the leading order, colour singlet component. Therefore,
the transition of a hard gluon into a \jpsi, via coupling to the $^3S_1$ colour
octet component of the \jpsi\ wave function, would be of order $v^4/\as^2$
relative to the standard fragmentation function of the CSM. A detailed
evaluation of the transition amplitudes, \cite{bf}, shows that the ratio of the
two contributions is actually $\sim 25 \pi^2  {\cal O}(v^4)/\as^2$, a number
large enough to explain the factor of 50 discrepancy between the data and the
predictions of the CSM model.

\begin{figure}[t]
\centerline{\psfig{figure=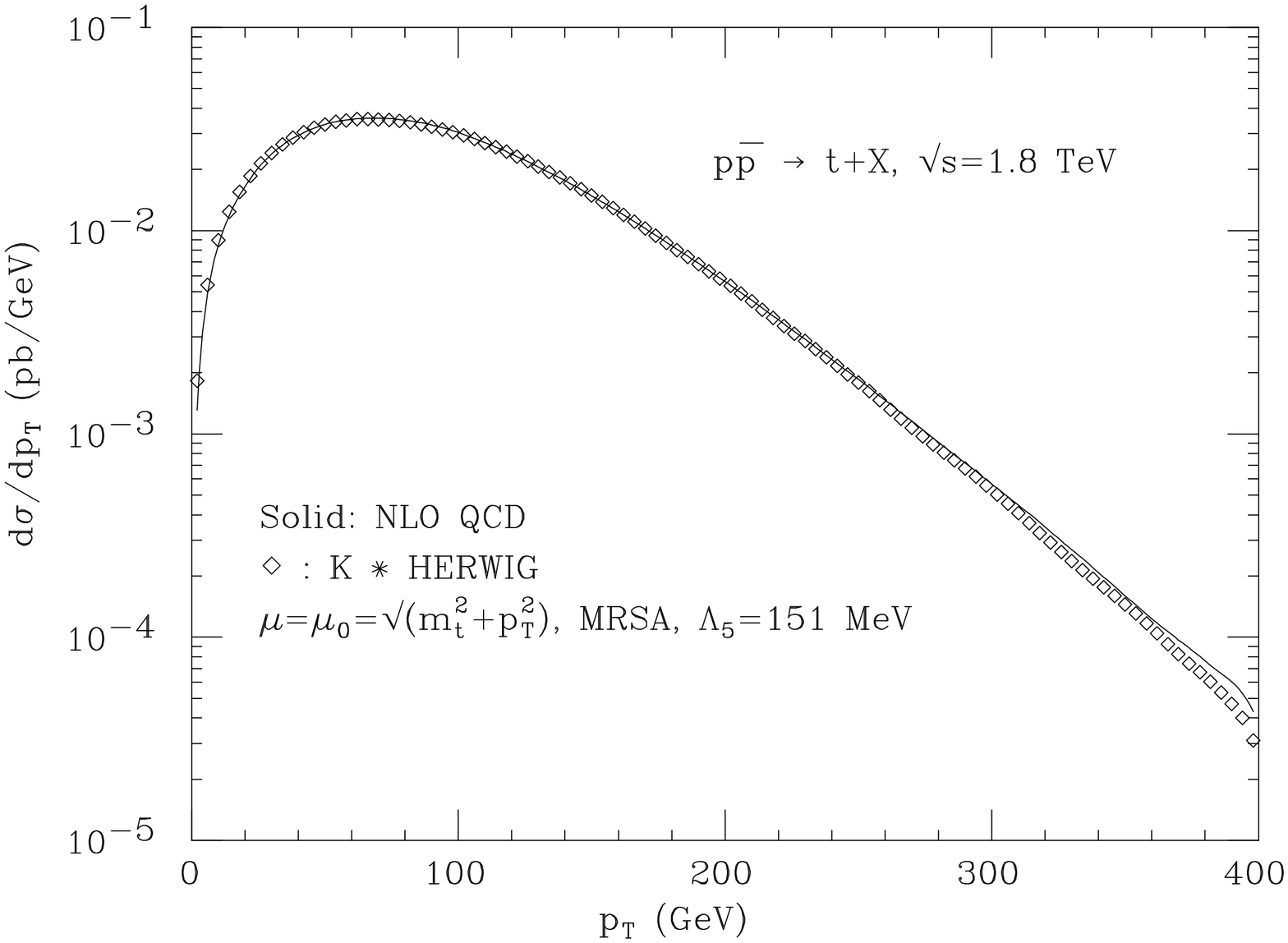,width=10cm,angle=90}}
\caption[]{ \label{toppt}
Inclusive \pt\ distribution of the top quark.
}
\end{figure}
The contribution of this colour octet production to the \psip\ distribution is
shown as a solid line in fig.~\ref{fpspdata}. Here the new non-perturbative
parameter \vevpsp, \ie the value of the overlap squared between
the $c\bar c$ colour octet state from gluon splitting and the \psip\ wave
function, was derived from a fit to the data. Its numerical value
has the right order of magnitude expected from the
$v^2$ suppression, consistently with what was suggested earlier. Similar
results can be obtained for the production  of \jpsi's \cite{cgmp}.

The model of quarkonium production via the colour octet mechanisms is now the
subject of intense work, and we will soon have available a more
complete picture of its implications, including the phenomenology of
$\Upsilon$ and charmonium in fixed target experiments.

\section{Top quark production}
\begin{figure}[t]
\centerline{\psfig{figure=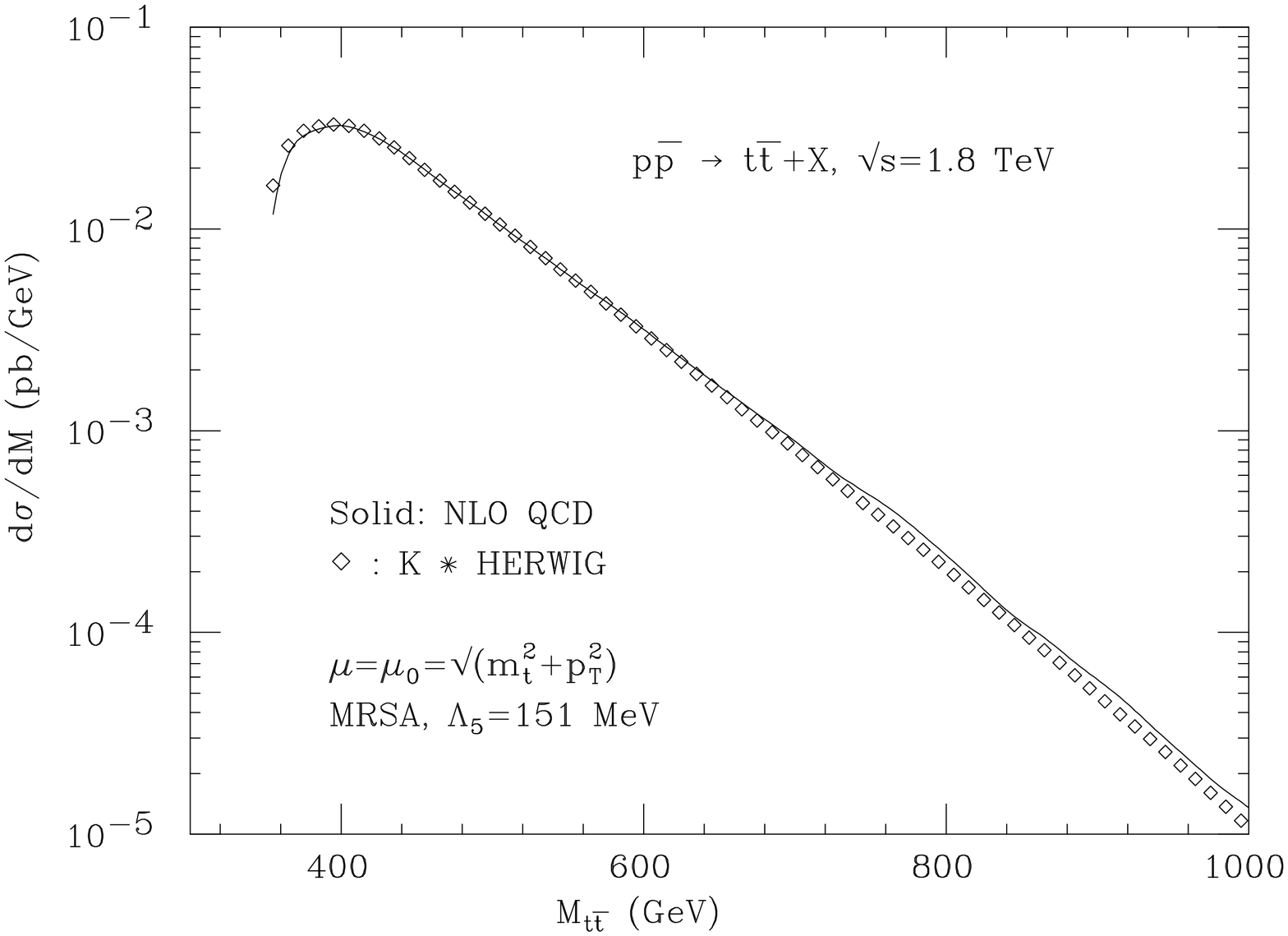,width=10cm,angle=90}}
\caption{ \label{tmass}
Invariant mass distribution of the $t\bar{t}$ pair.
}
\end{figure}
Now that the existence of the $top$ quark has been firmly established
via its detection in hadronic collisions \cite{toptev}, experimental
studies will focus on the determination of its properties. In particular, the
measurement of its mass and of the production cross section and distributions
will certainly be among the first studies of interest.
The production properties, should they display anomalies, could point to
the existence of exotic phenomena \cite{lane-et-al}. We present here some
kinematical distributions \cite{fmnr-top}\  that are of potential interest for
these comparisons.

The inclusive \pt\ distribution is sensitive to channels such as $W g \to t
\bar b$ \cite{wg}, which are found to contribute with a small cross section,
predominantly at low \pt.
The invariant mass of the pair is an obvious probe of the possible existence of
strongly coupled exotic resonances, such as technimesons \cite{lane-et-al}.
The transverse momentum of the pair is an indication of
the emission of hard hadronic jets in addition to those coming from the top
decays. The presence of these additional jets generates potentially large
combinatorial backgrounds to the reconstruction of the top mass peak
from the decay products \cite{cdf1}.
An accurate understanding of these backgrounds is
very important for a precise measurement of the top mass.

All the results we show were obtained using the NLO QCD matrix
elements \cite{bpt,mnr}, MRSA parton densities \cite{mrsa}, $m_{top}=176$ GeV.
Since most experimental studies are performed using shower Monte Carlo event
generators to simulate the behaviour of the events in the
detectors, we will compare our NLO results with those we obtained with HERWIG
\cite{herwig}. This is important in order to assess the reliability of the
theoretical inputs  used by the experiments.
Throughout our plots, we rescale the HERWIG calculations by the perturbative
$K$ factor given by the ratio of the NLO to LO results. The $K$
factor is of the order of 1.3 for all choices of parameters.

\begin{figure}[t]
\centerline{\psfig{figure=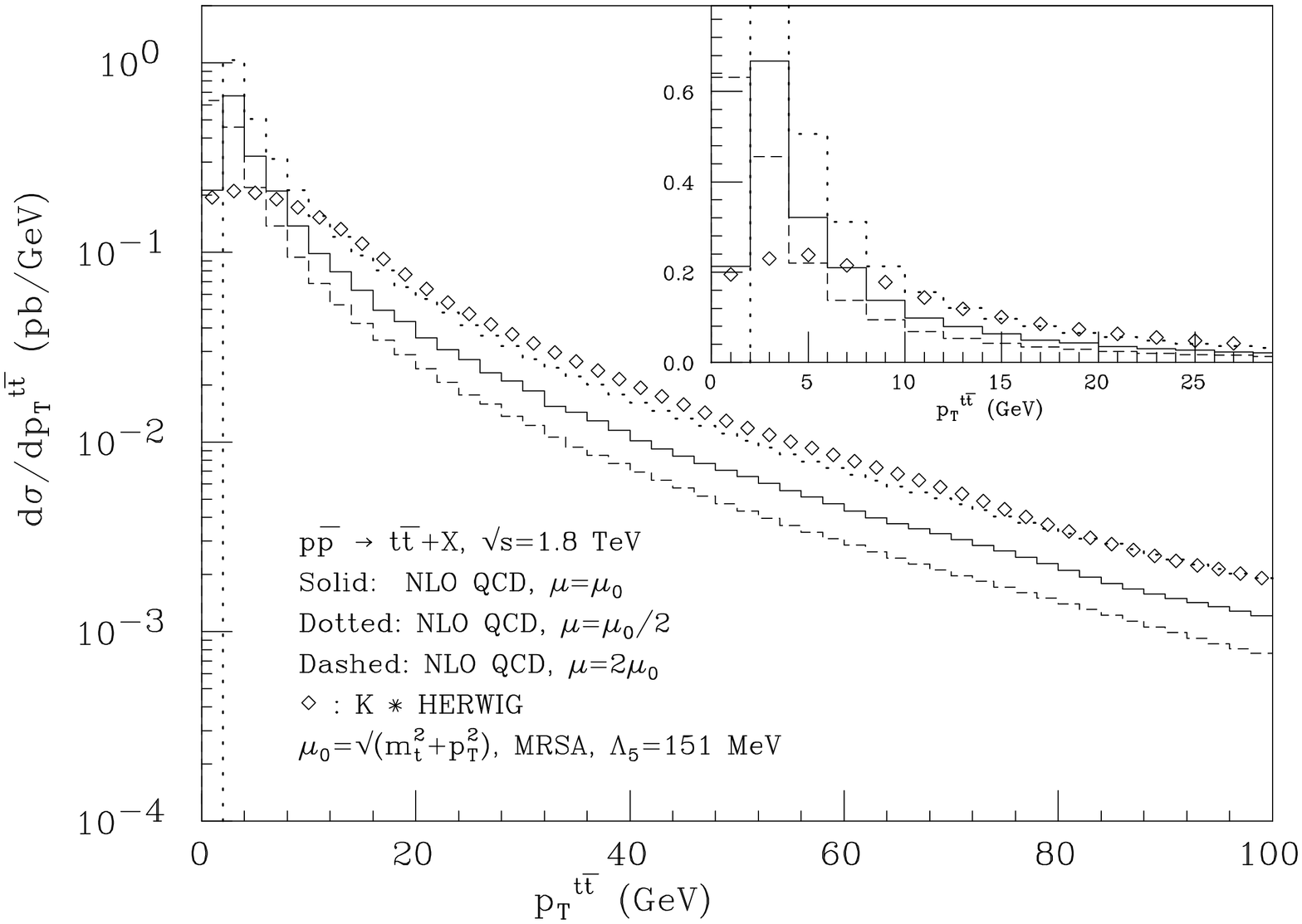,width=10cm,angle=90}}
\caption[]{ \label{fptpair}
Transverse momentum distribution of the $t\bar{t}$ pair.
}
\end{figure}
As an example of a single inclusive quantity, we show the top \pt\ distribution
in fig.~\ref{toppt}. The solid line corresponds to the NLO result obtained
using $\mur=\muf=\muo$. The square points correspond to the HERWIG prediction,
rescaled  by a $K$ factor equal to 1.34. The curves obtained by changing \mur\
and \muf\ by a factor 0.5 to 2 show an overall normalization change by
approximately $\pm 10\%$. No change in the shape is observed. The agreement
between the NLO and HERWIG results, and the stability under scale changes,
indicate that the prediction for the \pt\ distribution of top quarks is solid.
Aside from an overall small change in normalization, this result is not
affected by the inclusion of higher order soft gluon emission, \cite{smith}.

Similar conclusions \cite{fmnr-top}\ can be drawn for the rapidity
distribution, and for the distribution in invariant mass of the top pair, shown
in fig.~\ref{tmass}.

Those distributions which are trivial at leading order,
\dphi\ and \ptpair, are on the contrary most
sensitive to multiple gluon emission from the initial state.
This is because even small perturbations
can smear a distribution that at leading order is represented by a delta
function, as is the case for the \ptpair\ and \dphi\ ones.
The largest effect is
observed in \ptpair, fig.~\ref{fptpair}, where
we include the NLO curves relative to the
three choices of scales, $\mur=\muf=\muo$ (solid), \muo/2 (dots) and 2\muo
(dashes). Contrary to the previous cases, significant differences in shape
arise here among the three choices in the small-\pt\ region.
The HERWIG result (normalized to the area of the solid curve) is also shown.
The NLO and the HERWIG distributions assume the same shape only
for \ptpair\ larger than approximately 20 GeV.
We conclude that an accurate
description of the region \ptpair$<20$ GeV requires the resummation of leading
soft and collinear logarithms, as implemented in appropriate shower
Monte Carlo programs.

Studies of the angular correlations between bremstrahlung gluon jets and the
$b$-jets from the decay of the top quarks have been performed in
ref.~\cite{orr}. These authors found some important discrepancies between the
results obtained from a fixed order perturbative calculation and from HERWIG. A
full understanding of the origin of these discrepancies and a detailed study
of their possible impact on the combinatorial background to the reconstruction
of the top mass are in progress.

\end{document}